\documentclass[aps,twocolumn,prb,showpacs,10pt,floatfix,groupedaddress]{revtex4-1}
\usepackage{amssymb}
\usepackage{amsmath}
\usepackage{graphicx}
\usepackage{dcolumn}
\usepackage{bm}
\usepackage{textcomp}
\usepackage{color}
\usepackage{amsthm}
\usepackage{url}

\begin{document}

\title{Spectral butterfly and electronic localization in rippled-graphene nanorribons: mapping onto effective one-dimensional chains}

\author{Pedro Roman-Taboada$^{1}$}  
\email{peter89@fisica.unam.mx}
\author{Gerardo G. Naumis$^{2}$,$^{3}$}

\affiliation{1. Departamento de F\'{i}sica-Qu\'{i}mica, Instituto de
F\'{i}sica, Universidad Nacional Aut\'{o}noma de M\'{e}xico (UNAM),
Apartado Postal 20-364, 01000 M\'{e}xico, Distrito Federal,
M\'{e}xico}
\affiliation{2. School of Physics Astronomy and Computational Sciences, George Mason University, Fairfax, Virginia 22030, USA }
\affiliation{3. On sabbatical leave from Departamento de F\'{i}sica-Qu\'{i}mica, Instituto de
F\'{i}sica, Universidad Nacional Aut\'{o}noma de M\'{e}xico (UNAM),
Apartado Postal 20-364, 01000 M\'{e}xico, Distrito Federal,
M\'{e}xico}


\begin{abstract}
We report an exact map into one dimensional effective chains, of the tight-binding Hamiltonian for electrons in armchair and zigzag graphene nanoribbons with any uniaxial ripple. This mapping is used for studying the effect of uniaxial periodic ripples, taking into account the relative orientation changes between $\pi$ orbitals. Such effects are important for short wavelength ripples, while for long-wave ones, the system behaves nearly as strained graphene. The spectrum has a complex nature,
akin to the Hofstadter butterfly with a rich localization behavior. Gaps at the Fermi level and dispersionless bands were observed, as well. 
The complex features of the spectrum arise as a consequence of the quasiperiodic or periodic nature of the effective one dimensional system. Some features of these systems are understandable by considering  weakly coupled {dimers}. The eigenenergies of such {dimers} are highly degenerated and the net effect of the ripple can be seen as a perturbation potential that splits the energy spectrum. Several particular cases were  analytically solved to understand such feature. 
\end{abstract} 


\maketitle

%
\section{Introduction}
Graphene, which is a two dimensional (2D) crystal made from Carbon, has incredible electronic, optical, and mechanical properties \cite{Novoselov04, Lee08}. However, it is very difficult to grow perfectly flat graphene \cite{meyer07, meyersolid07, stolyarova07, Vinogradov12}. Instead, graphene presents corrugations and ripples that can improve or diminish its electronic conductivity \cite{Pereira09, guinea12, Pereira13, Olivaany14}. Hence, the understanding of how corrugations and ripples modify the electronic properties of graphene is a very important issue. Furthermore, such knowledge can provide a way for tailoring the electronic properties of graphene via mechanical deformation \cite{PereiraLetter09,Ong12,zhan12}. Even though the uniform and homogeneous strain has reached a level of good understanding \cite{Pereira09,Prada10,Maurice}, out of the plane deformation effects are not well understood. Moreover, most of the available theories are limited to the case of low energies or long wavelengths, in which it is possible to write an effective Dirac equation with effective pseudomagnetic fields \cite{Fogler10, Gradinar13, Carrillo14}. 

In a recent set of papers, we have shown that such picture can be modified for shorter wavelengths since a quasiperiodic fractal behavior, non-treatable under perturbation theory can appear \cite{Nosotros14,Nosotros214}. At the same time, an experimental observation of such  effect has recently been made \cite{Novoselov14}.  
{ This behavior is not new in graphene, actually, this fractality has been extensively studied in graphene under magnetic fields within 
the Dirac approach\cite{Nemec07,Chizhova14, Gumbs14, Yu14}, however, the study of this behavior as a consequence of strain or corrugation using a 
tight-binding approach has not been done.}
The aim of this work is to understand how ripples modify the electronic properties of armchair (AGN) and zigzag (ZGN)graphene nanoribbons.
To get such understanding, here we propose the study of uniaxial ripples using a tight-binding Hamiltonian. Uniaxial ripples already show the expected effects in more general cases, and at the same time, it is
possible to map the system into one-dimensional (1D) chains. This procedure is similar to that used for studying electrons in lattices under magnfetic fields, in which it is possible
to obtain the spectrum by studying the one dimensional Harper equation \cite{Harper55, Hofstadter76}. 
Once this connection is established, we propose the study of the physical effects using uniaxial periodic ripples. As we will see, the energy spectrum has a fractal structure with gaps at the Fermi level. This highlights the importance of the rational or irrational nature of the ripple's wavelength. Furthermore, we are able to solve analytically several cases, leading to expressions for the bands as a function of the ripple's parameters: wavelength and amplitude.

The layout of this work is the following: In Section \ref{Mapping}, we discuss the details of mapping AGNs and ZGNs under any uniaxial ripple into effective 1D chains. In Section \ref{uniperrip}, we study a particular case, uniaxial periodic ripple, using the previous maps. The properties of the energy spectrum as a function of the frequency of the ripple, the band structure, and the density of states (DOS) are discussed, as well. Finally, in Section \ref{conc}, our conclusions are presented.
\section{Mapping of uniaxial rippled graphene into an effective one-dimensional system}
\label{Mapping}
In this section, we will show how to reduce the study of uniaxial ripples in graphene into a
effective one-dimensional system. 
We start with a graphene's nanoribbon, as shown in Figs. \ref{arm} and \ref{zig}, with an
uniaxial deformation in the $y$-direction due to a ripple in the graphene's sheet. 
The new positions of Carbon atoms in the rippled graphene are
\begin{equation}
\bm{r}'=(\bm{r},z(\bm{r}))
\label{position}
\end{equation}
where $\bm{r}=(x,y,0)$ are the unrippled coordinates of Carbon atoms, and $z(\bm{r})$ is the height variable in terms of the position $\bm{r}$. 
To obtain the electronic properties, we use a one orbital next-nearest neighbor
tight binding Hamiltonian in a honeycomb lattice, 
given by \cite{Lu12},
\begin{equation}
H=-\sum_{\bm{r}',n} t_{\bm{r}',\bm{r}'+\bm{\delta}_{n}'} c_{\bm{r}'}^{\dag}
c_{\bm{r}'+\bm{\delta}_{n}'}+\text{H.c.},
\label{Hamiltonian}
\end{equation}
where the sum over $\bm{r}'$ is taken for all sites of the deformed lattice. The vectors
$\delta_{n}'$ point to the three next-nearest neighbor of
$\bm{r}'$. For unstrained graphene $\bm{\delta}_{n}'=\bm{\delta}_{n}$ where,
\begin{equation}
\bm{\delta}_{1}=\frac{a}{2}\left(1,-\sqrt{3},0\right)\ \
\bm{\delta}_{2}=\frac{a}{2}\left(1,\sqrt{3},0\right) \ \ \bm{\delta}_{3}=a(-1,0,0).
\end{equation}
and $c_{\bm{r}'}^{\dag}$, $c_{\bm{r}'}$ are the creation and annihilation
operators of an electron at the lattice position $\bm{r}'$.  
In this model, the hopping integral $t_{\bm{r}',\bm{r}'+\bm{\delta}_{n}'}$ depends upon the strain, since the overlap
between graphene orbitals is modified as the inter-atomic distances change. When corrugation is presented, the orbitals $\pi$ are no longer parallel. Let us denote $\theta_{\bm{r}'}$ to the angle which determines the relative orientation of a carbon  atom in the position $\bm{r}'$ in the graphene nanoribbon. This angle depends on the local curvature of the layer.
The effect of the relative orientation change of the $\pi$ orbitals and the inter-atomic distances changes are described by \cite{Eun08,Neto09,Guinea10}
\begin{equation}
\begin{split}
t_{\bm{r}',\bm{r}'+\bm{\delta}_{n}'}=& t_0 \left[1+\alpha\left(1-\bm{N}_{\bm{r}'}\cdot\bm{N}_{\bm{r'+\bm{\delta}_n'}}\right)\right]\\
&\times\exp{\left[-\beta(l_{\bm{r}',\bm{r}'+\bm{\delta}_n'}/a-1)\right]},
\end{split}
\end{equation}
where $\bm{N}_{\bm{r}'}$ is the unit vector normal to the surface in the site $\bm{r}'$,  given by,
\begin{equation}
\bm{N}_{\bm{r}'}=\frac{\hat{\bm{z}}-\nabla z}{\sqrt{1+(\nabla z)^2}},
\label{N}
\end{equation}
and $\nabla=(\partial_x,\partial_y)$ is the 2D gradient operator while $\hat{\bm{z}}$ is the unit vector in the perpendicular direction to the plane.
 $l_{\bm{r}',\bm{r}'+\bm{\delta}_n'}$ is the interatomic distance between two neighbors sites after a ripple is applied, and $\alpha\approx0.4$ is a constant that takes account for the change of the relative orientation of the $\pi$ orbitals. Here
$\beta\approx 3.37$, and $t_0\approx 2.7\,eV$ corresponds to graphene without strain.
{ The unrippled bond length is denoted by $a$. For the propose of this paper, it is natural to measure the distances in units of $a$,
which is equivalent to set $a=1$. In other words, all the distances and lengths will be measured in units of $a$}.
It is important to remark that  contributions from terms containing $\beta$ are due to
distance changes, while terms dependent on $\alpha$ account for angular changes of the orbital
overlap. As we will see, $\beta$-depend terms tend to shrink 
the energy spectrum whereas $\alpha$-depend terms tend to stretch it.

Now, for uniaxial ripple the interatomic distances between carbon atoms can be written as
\begin{equation}
\begin{split}
l_{\bm{r}',\bm{r}'+\bm{\delta}_n'}=&\left|\left|\bm{\delta}_n+\left[z\left(y+\delta^{(y)}_n\right)-z(y)\right]\,\hat{\bm{z}}\right|\right|\\
=&\sqrt{1+\left[z\left(y+\delta^{(y)}_n\right)-z(y)\right]^2}.
\end{split}
\end{equation}
Recently, we have shown that is possible to map armchair and zigzag graphene nanoribbon (AGNs and ZGNs) under uniaxial strain onto an effective one dimensional system\cite{Nosotros14,Nosotros214}. In the next  subsection, we extend such results for AGN and ZGN under uniaxial ripples.

Before entering the details of the maps and for comparison purposes with other works, it is important to remark that for small amplitude and long wavelength ripples, we have $\left(1-\bm{N}_{\bm{r}'}\cdot\bm{N}_{\bm{r'+\bm{\delta}_n'}}\right) \approx \theta_{\bm{r}',\bm{r}'+\bm{\delta}_n'}^2/2$, where $\theta_{\bm{r}',\bm{r}'+\bm{\delta}_n'}$ is the angle between $\pi$ orbitals in the sites $\bm{r}'$ and $\bm{r}'+\bm{\delta}_n'$. In such a case, since $\alpha<\beta$  and the angle correction is quadratic, it follows that,
\begin{equation}
\begin{split}
t_{\bm{r}',\bm{r}'+\bm{\delta}_{n}'}&\approx t_0 \left[1+\frac{\alpha}{2}\theta_{\bm{r}',\bm{r}'+\bm{\delta}_n'}^2\right]
\exp{\left[-\beta(l_{\bm{r}',\bm{r}'+\bm{\delta}_n'}-1)\right]}\\
&\approx t_0 \left[1-\beta\left(l_{\bm{r}',\bm{r}'+\bm{\delta}_n'}-1\right)\right].
\end{split}
\end{equation}
In this limit, the model resembles graphene's nanoribbons under planar strain. As we will show, our general computations
are consistent with this limit, providing a test for the method presented here.
\subsection{Armchair graphene nanoribbon}
\begin{figure}
\includegraphics[scale=0.325]{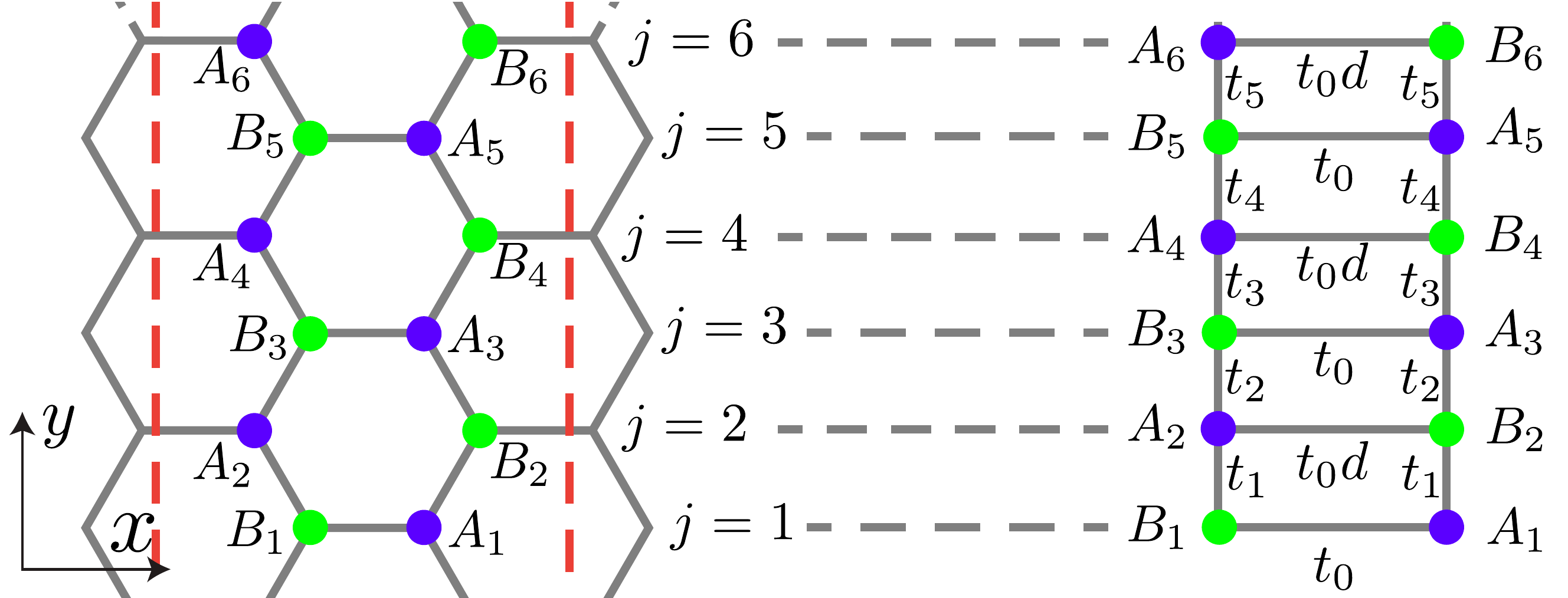}
\caption{\label{arm} (Color online) Primitive cell for AGNs (on the left between the red dotted lines) which is made for $2N$ atoms, i. e., $N$ atoms of type $A$ (blue solid circles) and $N$ atoms of type $B$ (green solid circles), we order the basis as $A_1$, $B_2$, ..., $B_{N-1}$, $A_N$, $B_1$, $A_2$, ..., $B_{N-1}$, $B_N$. We vary the height of each atom along the $y$-direction. For this case the hopping parameter just depends upon the $y$-component of the in-plane atom's positions. The system is equivalent to the one-dimensional effective ladder shown at the right, where the label $j$ corresponds to each ladder step in the $y$-direction, $t_j$ is the hopping integral for hopping from the atom $A_{j-1}$ ($B_{j-1}$) to the atom $B_{j}$ ($A_j$), and $d$ is a coefficient that depend upon the momentum $k_x$.}
\end{figure}
When an uniaxial ripple in the $y$-direction is applied, it is possible to describe the electronic properties of the AGN by an effective one-dimensional Hamiltonian. We start by labeling the atom's positions as shown in Fig. \ref{arm}, i. e., we order the basis as $A_1$, $B_2$, $A_3$, ..., $A_{N-1}$, $B_N$, and $B_1$, $A_2$, $B_3$, ..., $A_{N-1}$, $B_N$. Thus, the effective one-dimensional Hamiltonian can be written as\cite{Nosotros214}
 \begin{equation}
\begin{split}
H_{AGN}(k_x)=&\sum_jt_0\left[d(k_x)\,a_{2j}^{\dag}b_{2j}+a_{2j+1}^{\dag}b_{2j+1}\right]\\
&+\sum_j\, t_j^{\text{AGN}}a_{j}^{\dag}b_{j+1} +\text{H.C.},
\end{split}
\end{equation}
where $d(k_x)=\exp{(ik_x)}$, $a_j$, $a_j^{\dag}$, and $b_j$, $b_j^{\dag}$ are the annihilation and creation operators in the sublattices $A$ and $B$ in graphene, respectively. This effective Hamiltonian describes two modulated chains, as is shown in the Fig. \ref{arm}. $t_j^{\text{AGN}}$ is the hopping parameter between the $j+1$ and $j$ atoms in the $y$ direction, given by,
\begin{equation}
\begin{split}
t_j^{\text{AGN}}=&t_0\left[1+\alpha\left(1-\bm{N}_{j+1}^{\text{AGN}}\cdot\bm{N}_j^{\text{AGN}}\right)\right]\times\\
&\exp{\left[-\beta(l_{j+1,j}^{\text{AGN}}-1)\right]}
\end{split}
\end{equation}
where
\begin{equation}
l_{j+1,j}^{\text{AGN}}=\sqrt{1+\left[z(y_{j+1}^{\text{AGN}})-z(y_j^{\text{AGN}})\right]^2},
\end{equation}
is the interatomic distance between the atoms in sites $j+1$ and $j$, and
\begin{equation}
y_j^{\text{AGN}}=y_A^{\text{AGN}}(j)=y_B^{\text{AGN}}(j)=\sqrt{3}(j-1)/2,
\end{equation}
are the positions for atoms $A$ and $B$, and $j=1$, $2$, ..., $N$, labels the sites along $y$-direction for pristine graphene. Finally the unitary normal vector is $\bm{N}_j^{\text{AGN}}=\bm{N}(y_j^{\text{AGN}})$ defined as in Ec. (\ref{N}).
\subsection{Zigzag graphene nanoribbon}
\begin{figure}
\includegraphics[scale=0.44]{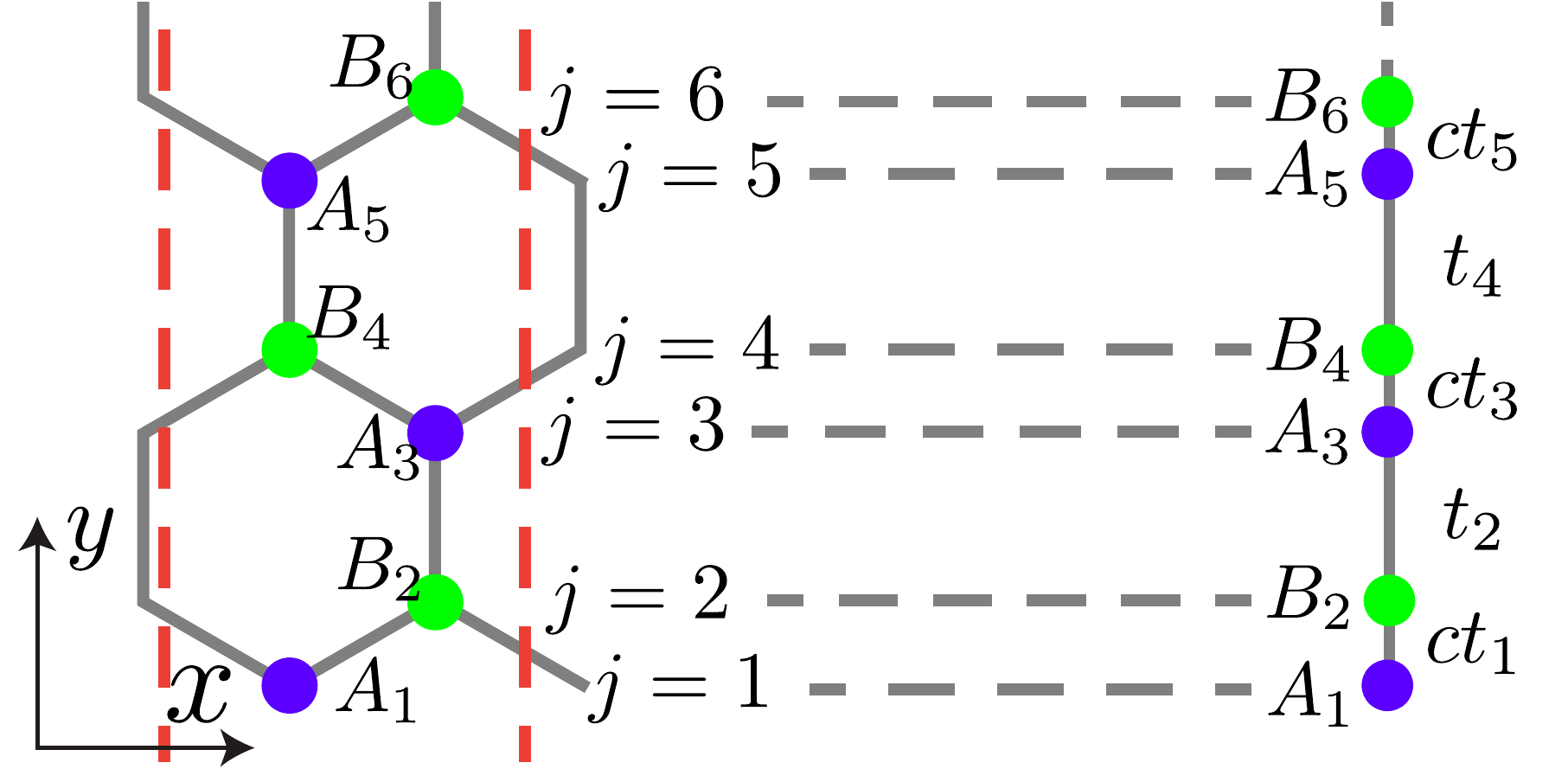}
\caption{\label{zig} (Color online) Primitive cell for ZGNs (to the left, delimited by two red dotted lines) made of $N$ atoms ($N/2$ atoms of type $A$ and $N/2$ atoms of type $B$). The basis is ordered as following $A_1$, $B_2$, ..., $A_{N-1}$, $B_N$. The height $z$ of every atom is modified only in the $y$-direction. Then, the hopping parameter is no longer constant but depends upon the $y$-component of the atom's positions. This system can be mapped into a one-dimensional effective chain (on the left) where $j$ labels the site along the zigzag direction, $t_j$ is the hopping parameter for hopping from the $(j-1)$-th atom to the $j$-th atom, and $c$ is a coefficient that depends on the momentum in the $x$-direction.}
\end{figure}
Similarly, when we apply an uniaxial ripple to a zigzag graphene nanoribbon (ZGN), it is possible to map the system into a one-dimensional effective chain. If we label the basis as $A_1$, $B_2$, ..., $A_{N-1}$, $B_N$ as shown in Fig. \ref{zig}, the resulting Hamiltonian is\cite{Nosotros14}
\begin{equation}
\begin{split}
H_{ZGN}(k_x)=&\sum_{j} \left[ c(k_x)t_{2j+1}^{\text{ZGN}} a_{2j+1}^{\dag} b_{2j+2}+t_{2j+2}^{\text{ZGN}} b_{2j+2} a_{2j+3}^{\dag} \right]\\&+\text{H.C.},
\end{split}
\end{equation}
where $c(k_x)=2\cos{\sqrt{3}k_x/2}$ and 
 \begin{equation}
\begin{split}
t_j^{\text{ZGN}}=&t_0\left[1+\alpha\left(1-\bm{N}_{j+1}^{\text{ZGN}}\cdot\bm{N}_j^{\text{ZGN}}\right)\right]\times\\
&\exp{\left[-\beta(l_{j+1,j}^{\text{ZGN}}-1)\right]},
\end{split}
\end{equation}
 is the hopping parameter between the sites $j+1$ and $j$ in the $y$-direction, $\bm{N}_j^{\text{ZGN}}=\bm{N}(y_j^{\text{ZGN}})$ defined as in Ec. (\ref{N}), and
\begin{equation}
l_{j+1,j}^{\text{ZGN}}=\sqrt{1+\left[z\left(y_{j+1}^{\text{ZGN}}\right)-z\left(y_j^{\text{ZGN}}\right)\right]^2}
\end{equation}
where
\begin{equation}
y_{j}^{\text{ZGN}}=y^{\text{ZGN}}(j)=\frac{1}{4}\left\{3j+\frac{1}{2}\left[1-(-1)^j\right]\right\},
\end{equation}
are the positions of carbon atoms in unrippled graphene and $j=1$, $2$, ..., $N$ labels the sites as is displayed in Fig. \ref{zig}.
\section{Uniaxial periodic ripple}
\label{uniperrip}
Let us now study in this section the particular case of a periodic uniaxial ripple. This kind of corrugation is commonly observed when graphene is grown on a substrate\cite{Vinogradov12}. In particular, we will consider that the periodic uniaxial ripple has the following form
\begin{equation}
z(y)=\lambda \cos{(2\pi \sigma y+\phi)}.
\label{zr}
\end{equation}
{ This particular oscillation contains three parameters: wavelength (controlled by the parameter $\sigma$), amplitude (controlled by $\lambda$), and phase (controlled by $\phi$). Thus, $\sigma$ is translated into a ripple with a spatial wavelength $\Lambda$  such that $\Lambda=2\pi a/\sigma$. Small $\sigma$'s compared with the lattice parameter $a$ are translated into long-wavelength ripples. The amplitude $\lambda$ is the maximal height reached by the rippled, usually given in nanometers or in percentages of $a$. 

Now we wil discuss briefly the feasibility of such a particular ripple. Since graphene exhibits a high asymmetry in tensile versus compressive strain, i. e., while the C-C bond length can be tensile up to 25\%\cite{Lee08} of the lattice parameter, it is almost incompressible because this would induce out of plane deformations. Thus, in general, to produce this ripple it is enough to apply uniaxial strain. Also, it has been observed that growing graphene on a substrate 1D periodic graphene ripple can be built by using thermal strain engineering and the anisotropic stress due to the substrate\cite{Bai14}. By the other hand, as we will see later, we will use a big ripple's amplitude ($\lambda=$80\% of the lattice parameter) in all of our plots for illustrating proposes. Even though this value is high, the most important of our results only depend upon the ripple's wavelength ($\sigma$) and is valid for all values of $\lambda$}.

Another important aspect of the electronic properties is the wave function localization. For studying it, we will use the normalized participation ratio (NPR), defined as,
\begin{equation}
NPR (E)=\frac{\ln{\sum_{j=1}^N |\psi (j)|^4}}{\ln{N}}.
\end{equation}
This quantity is a measure of wavefunctions localization\cite{Naumis07}, for extended states, $NPR\rightarrow -1$ (blue color in the figures), whereas it tends to zero for localized states (red color in the figures).
In the next subsection, we will study the physical effects in the electronic properties of AGNs and ZGNs under the previous periodic ripples.
\subsection{AGN with uniaxial periodic ripples}
\begin{figure}
\includegraphics[scale=0.43]{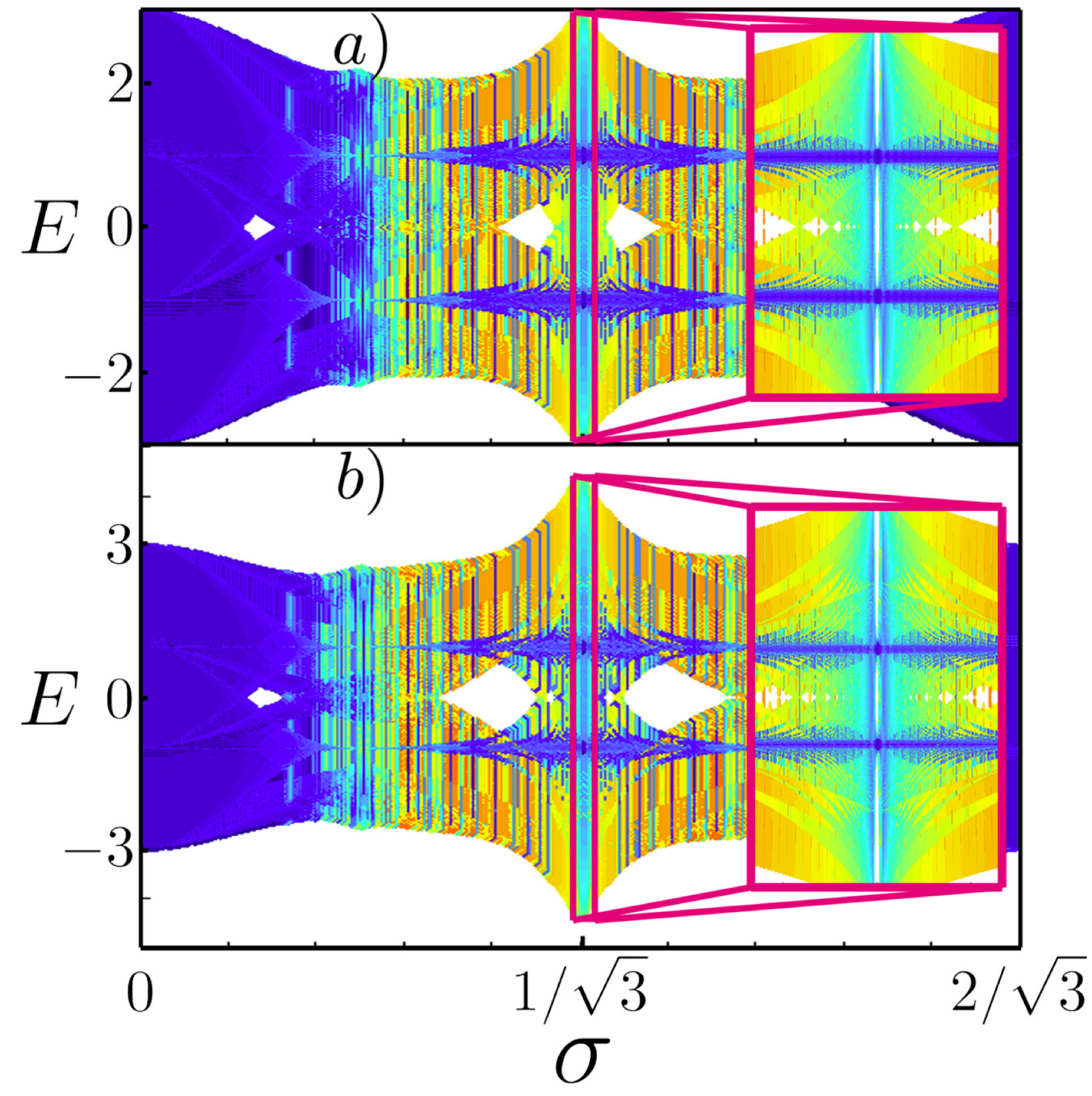}
\caption{\label{sigarm} (Color online) Energy spectrum of AGNs as a function of $\sigma$ for
$\lambda=0.8$, using $a)$ $\alpha=0.4$ and $b)$ $\alpha=0$ obtained by solving the Schr\"odinger
equation for a system of 400 atoms, using 250 grid points for sampling $k_x$ and
with ciclic boundary conditions. The different colors represents the localization
participation ratio $NPR(E)$.}
\end{figure}
When we apply an uniaxial periodic ripple given by Eq. (\ref{zr}) to AGNs, the hopping integral becomes
\begin{equation}
\begin{split}
&t_j^{\text{AGN}}=t_0\left[1+\alpha\left(1-\bm{N}_{j+1}^{\text{AGN}}\cdot\bm{N}_j^{\text{AGN}}\right)\right]\times\\
&\exp{\left\{-\beta\left[\sqrt{1+4\lambda^2\sin^2{\left(\frac{\sqrt{3}}{2}\pi\sigma\right)}\xi_{\text{A}}^2(j+1/2)}-1\right]\right\}}
\label{tagn}
\end{split}
\end{equation}
where $\xi_{\text{A}}(j)=\sin{(\sqrt{3}\pi\sigma j+\phi)}$. To get a better understanding, it is worth considering the small amplitude case. Using Eq. (\ref{tagn}), the hopping  inter chains parameter becomes
\begin{equation}
\begin{split}
t_j^{\text{AGN}}&\approx t_0\exp{\left[-2\beta\lambda^2\sin^2{\left(\frac{\sqrt{3}}{2}\pi\sigma\right)}\xi^2(j+1/2)\right]}\\
& \approx t_0-2t_0\beta\lambda^2\sin^2{\left(\frac{\sqrt{3}}{2}\pi\sigma\right)}\xi^2(j+1/2).
\end{split}
\end{equation}
This expression is quite similar to the hopping integral that appears in the off-diagonal Harper model\cite{Harper55}, the main difference here is that all terms are squared. Hence, we expect the period in $\sigma$ to be half of the period of AGNs under uniaxial periodic strain. Otherwise, both spectra would be really similar in the low energy or long wavelength limit, as can be confirmed in Fig. \ref{sigarm}. Therein, it is shown the spectrum of $H_{\text{AGN}}$ as a function of $\sigma$ for $a)$ $\alpha=0$ and $b)$ $\alpha=0.4$, obtained using cyclic boundary conditions by diagonalizing the resulting matrix for each value of $k_x$. Many interesting features are observed. First,  the spectrum has a fractal nature with gaps at the Fermi level. Second, we observe that localized states coexist with extended ones. It is easy to understand this feature, since when the period of the lattice is commensurable with the ripple's period the system behaves as a modulated crystal, and the states have a Bloch nature which are extended. When the periods are incommensurate between them, the system behaves as a quasicrystal and the wave functions tend to have different localization properties. Finally, we observe that the effect of relative orientation changes between $\pi$ orbitals (i.e., the $\alpha$-depend terms effects) is to produce a widening of the spectrum, which becomes important for $\sigma$ around $1/\sqrt{3}$. 
\begin{figure}
\includegraphics[scale=0.41]{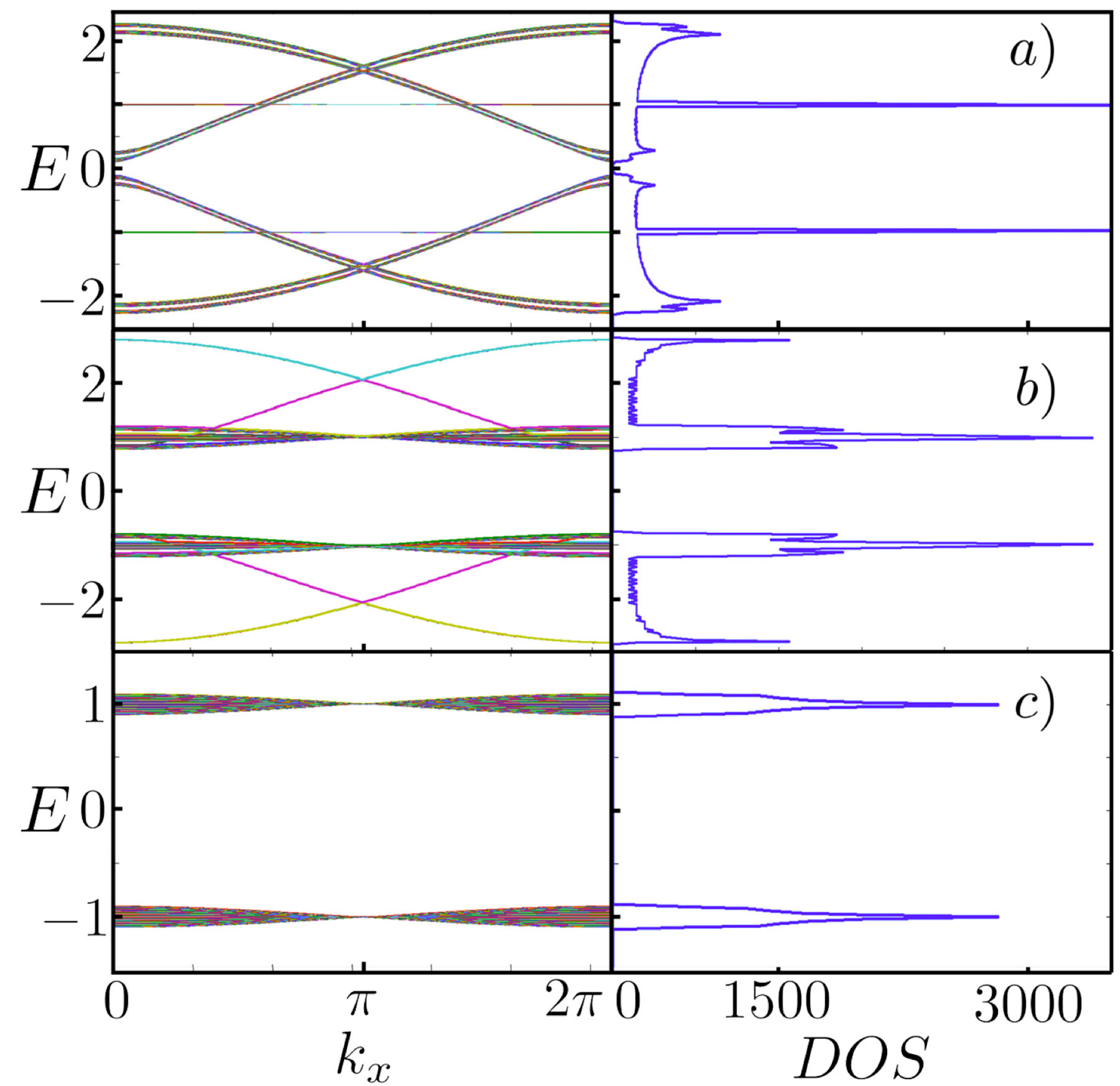}
\caption{\label{armEvskx} (Color online) Band structure and density of states ($DOS$) of an AGN for $\phi=0$ and $\lambda=0.8$ using $a)$ $\sigma=0.6/\sqrt{3}$, $b)$ $\sigma=0.8/\sqrt{3}$, and $c)$ $\sigma=1/\sqrt{3}$ for a system with 200 atoms. Cyclic boundary conditions were used.}
\end{figure}
To gain further insight on the spectrum, consider the interesting transition seen as $\sigma$ goes from zero to $\sigma=1/\sqrt 3$, as shown in detail on Fig. \ref{armEvskx}. Therein, the band structure is displayed for $a)$ $\sigma=0.6/\sqrt{3}$, $b)$ $\sigma=0.8/\sqrt{3}$, and $c)$ $\sigma=1/\sqrt{3}$. This transition goes from unrippled graphene to a system made by weak coupled { dimers}. The { dimers}
are made from pairs of sites joined by an horizontal bond in Fig. \ref{arm}. The { dimers} appear  since for $\sigma=1/\sqrt{3}$, the hopping parameter becomes $t_j^{\text{AGN}}=t(\lambda=0.8)\approx0.05t_0$. The exact expression of $t(\lambda)$ will be given in the next subsection. Thus, $t(\lambda=0.8)<<t_0$ and $t_j$ can be considered as a weak perturbation to a system made by pure { dimers}. The eigenenergies of the { dimers} are highly degenerate, with values $E=\pm t_0$ as observed in Fig. \ref{armEvskx} a). The effect of $t(\lambda)$ is just a widening  around these values, giving an spectrum in the intervals $[\pm t_0-t(\lambda=0.8),\pm t_0+t(\lambda=0.8)]$, as observed in Fig. \ref{armEvskx} a). As $\sigma \rightarrow 0$, the { dimers} evolve into the Van Hove singularity at $E=\pm t_0$ observed in unrippled graphene. Also, the system can be treated as a ladder with $t_j^{\text{AGN}}=\left<t\right>+\delta_j$, 
where $\left<t\right>$ is the average hopping parameter, and $\delta_j$ is a small perturbing potential $\delta_j << \left<t\right>$. For example, the case c) in Fig. \ref{armEvskx}, corresponds to  weakly coupled squares.

\begin{figure}
\includegraphics[scale=0.32]{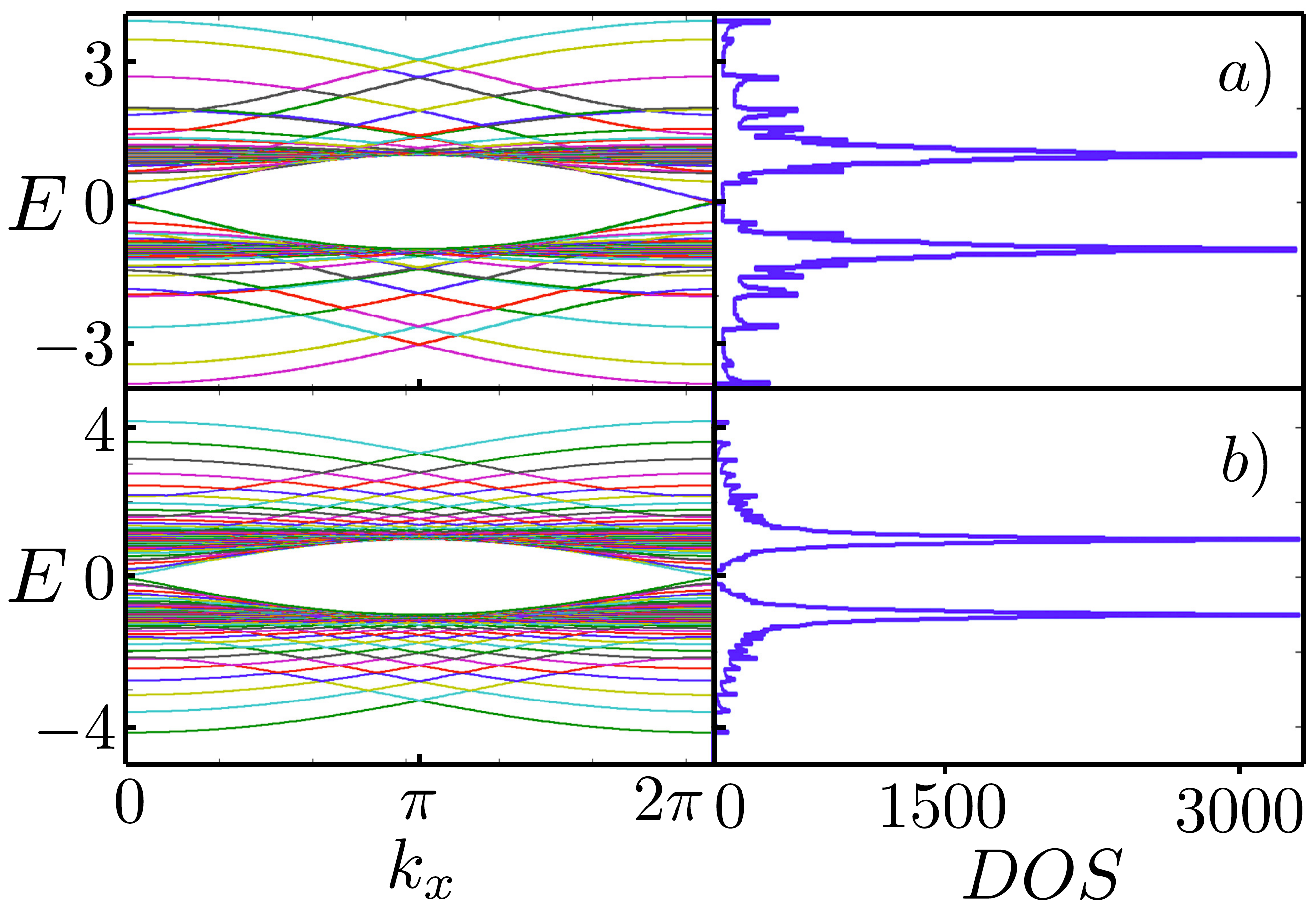}
\caption{\label{armEvskxinc} (Color online) Band structure and DOS for an AGN using $a)$ $\sigma=4\sqrt{5/3}$ and $b)$ $\sigma=0.4\sqrt{7/3}$ for $\alpha=0.4$ and the same conditions as in Fig. \ref{armEvskx}. Note that there are two partially flat bands at $E=\pm1$ and that the DOS is spiky.}
\end{figure}

Before showing how the case $\sigma=1/\sqrt 3$ can be solved analytically leading to weakly coupled { dimers}, let us discuss the band structures displayed in Fig. \ref{armEvskxinc}. Therein, it is shown the band structure and the DOS for $\sigma$ values that are incommensurable with the period of unrippled AGN's. In these cases, the { dimer}'s model is still useful. When the $t_j^{\text{AGN}}\rightarrow0$ the effective system is made of { dimers} with two different intra-{ dimer} hopping parameter,  $t_0$ and $t_0 d(k_x)$, with eigenenergies $\pm1$. These energies are highly degenerated because of the factor $d(k_x)$. When  $t_j^{\text{AGN}}$ becomes quasiperiodic, the degeneration is broken and the spectrum is fragmented, as  observed in Fig. \ref{armEvskxinc}. However, the other { dimers} with hopping parameter given by $t_0$ are not affected. This kind of { dimers} are responsible of the partly flat bands at $\pm t_0$.
\subsubsection{Particular case $\sigma=1/\sqrt{3}$}
For $\sigma=1/\sqrt{3}$, the eigenergies can be calculated exactly. At this particular $\sigma$ value the hopping integral can be written as follows,
\begin{equation}
t_j^{\text{AGN}}=t(\lambda)=t_0\exp{\left[-\beta\left(\sqrt{1+4\lambda^2}-1\right)\right]}.
\label{tlambda}
\end{equation}
Hence, the unit cell only contains four different kind of sites. Note that for big $\lambda$ we have $t(\lambda)\rightarrow0$ and the effective system is made of two different { dimers}, corresponding to horizontal pairs of atoms, with inter-{ dimer} hopping parameter $t_0$ and $t_0d(k_x)$. This confirms the previous discussion and the band structure shown in Fig. \ref{armEvskx} $c)$. The eigenenergies for such a system are,
\begin{equation}
E(k_x)=\pm t_0\sqrt{1+[t(\lambda)]^2\pm2t(\lambda)|\cos{(k_x/2)}|}.
\end{equation}
From the previous equation one can prove that the gap's size is
\begin{equation}
\Delta_{\text{AGN}}=2|t(\lambda)-t_0|.
\end{equation}
which is the same that we obtained using the intervals $[\pm t_0-t(\lambda=0.8),\pm t_0+t(\lambda=0.8)]$ when perturbation theory was applied. It is worth finding the minimum value of $\lambda$ for opening a gap. It is easy to show that this occurs for any $\lambda>0$, {  which is an important result for applications, due to the fact that is possible to tailor the gap's size at the Fermi level by using the previous equation with a ripple's amplitude within the elastic response of graphene.}
\subsection{ZGN with periodic uniaxial ripples}
\begin{figure}
\includegraphics[scale=0.41]{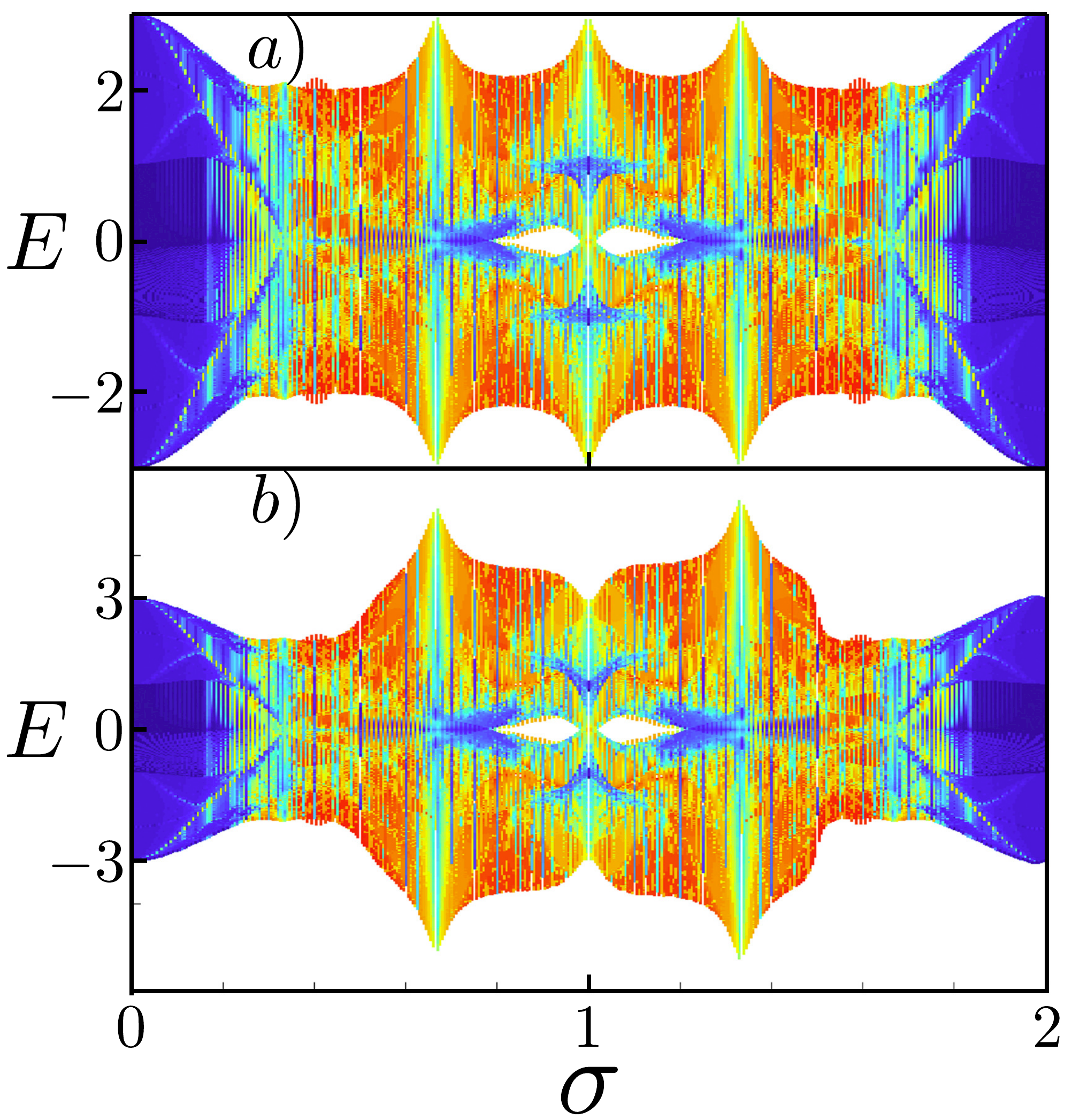}
\caption{\label{sigzig} (Color online) Energy spectrum of ZGN as function of $\sigma$ for
$a)$ $\lambda=0.8$ and $\alpha=0$, $b)$ $\lambda=0.8$ and $\alpha=0.4$ obtained by solving the Schr\"odinger
equation for a system of 80 atoms, using 250 grid points for sampling $k_x$ and
with ciclic boundary conditions. The different colors represents the localization
participation ratio $NPR(E)$. Note that the gaps at the Fermi level are less and smaller than in AGN rippled, the wavefunctions are more localized, as well.}
\end{figure}

When a ripple given by Eq. (\ref{zr}) is applied to ZGNS, the hopping integral becomes,
\begin{equation}
\begin{split}
&t_j^{\text{AGN}}=t_0\left[1+\alpha\left(1-\bm{N}_{j+1}^{\text{ZGN}}\cdot\bm{N}_j^{\text{ZGN}}\right)\right]\times\\
&\exp{\left\{-\beta\left[\sqrt{1+4\lambda^2\sin^2{\left(\frac{\pi}{2}\sigma\varphi_j\right)}\xi_{\text{Z}}^2(3j/2+1)}-1\right]\right\}},
\end{split}
\end{equation}
where $\varphi_j=[3+(-1)^j]/4$ and $\xi_{\text{Z}}(j)=\sin{(\pi\sigma j+\phi)}$. Further insight can be obtained by analyzing the case of small amplitude ripples, in which the hopping parameter is the following,
\begin{equation}
\begin{split}
t^{\text{ZGN}}_j&t_0\approx\exp{\left\{-2\beta\lambda^2\sin^2{\left(\frac{\pi}{2}\sigma\varphi_j\right)}\xi_{\text{Z}}^2(3j/2+1)\right\}}\\
&\approx t_0-2t_0\beta\lambda^2\sin^2{\left(\frac{\pi}{2}\sigma\varphi_j\right)}\xi_{\text{Z}}^2(3j/2+1).
\end{split}
\end{equation}
Again, this hopping integral is very similar to that for ZGNs under uniaxial periodic strain\cite{Nosotros14}. As in AGNs, the main difference is that all terms are squared. That makes the period to be a half of the ZGNs uniaxial periodic strained, as confirmed in Fig \ref{sigzig}.
\begin{figure}
\includegraphics[scale=0.411]{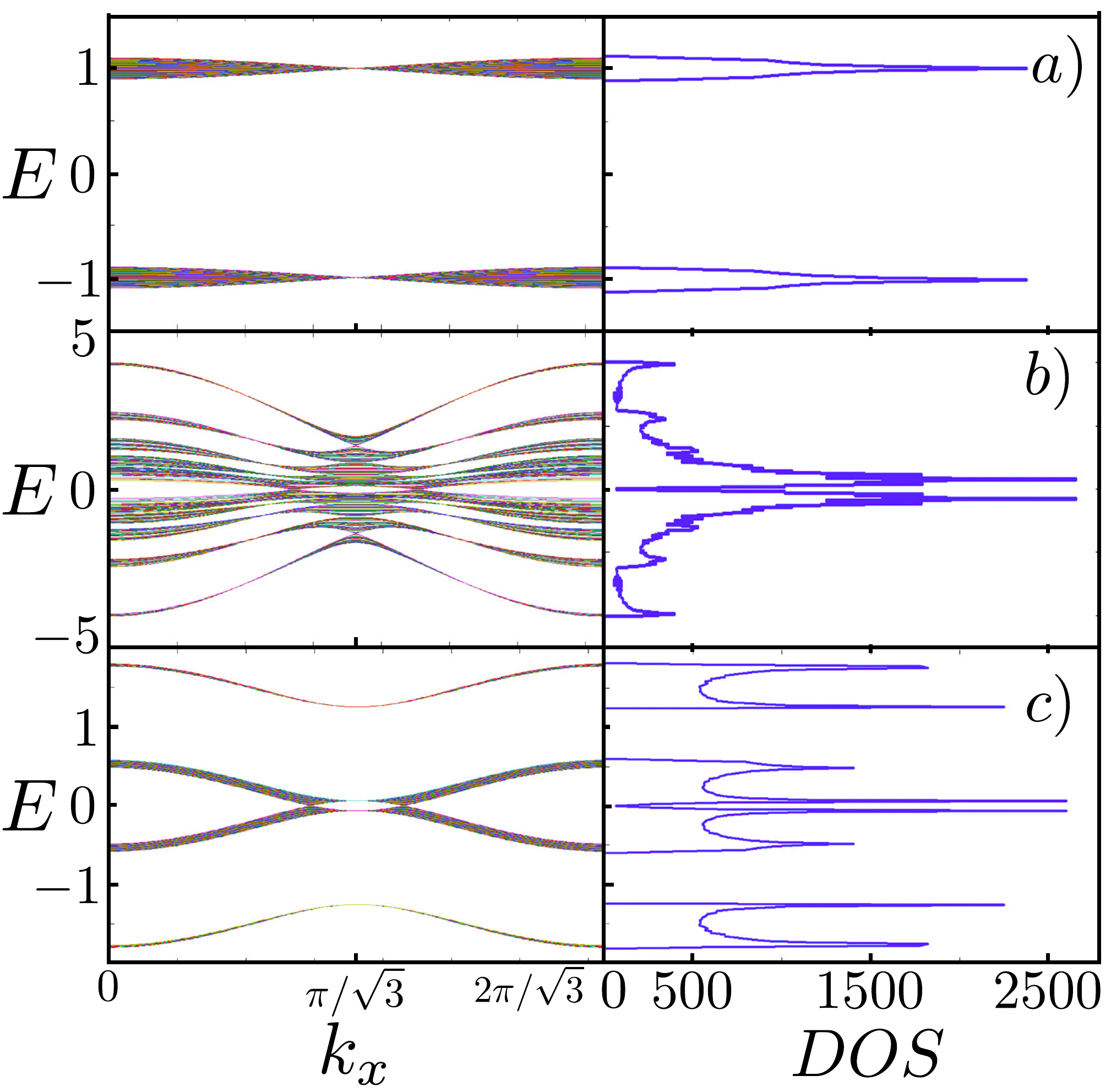}
\caption{\label{zigEvskx} (Color online) Band structure and density of states ($DOS$) of a ZGN for $\phi=0$, $\lambda=0.8$ using $a)$ $\sigma=1$, $b)$ $\sigma=(\sqrt{5}-1)/2$, and $c)$ $\sigma=1/4$. Same conditions as in Fig. \ref{sigzig} were used. Note the big gap opened at the Fermi level in $a)$, in $b)$ the band structure is fragmented, hence the $DOS$ has a lot of spikes, and in $c)$ the $DOS$ is similar to linear chains weakly interacting.}
\end{figure}
There in, we show the energy spectrum for ZGNs obtained using cyclic boundary conditions and by diagonalizing the corresponding matrix for each value of $k_x$. Two different values of $\alpha$ were used, $a)$ $\alpha=0$ and $b)$ $\alpha=0.4$.  There are many interesting features. First, the states are more localized than in the AGN case. Second, the gaps at the Fermi level are smaller than the gaps in AGNs. Third, ZGNs are more sensible to the effect of $\alpha$, however, as in AGNs, this effect is just a widening of the energy spectrum. The structure still being the same as in the case $\alpha=0$, specially for values near or at $\sigma=2/3,4/3$ (see Fig. \ref{sigzig} a) and b)). For $\sigma=2/3,4/3$, we have, 
\begin{equation}
\begin{split}
t_j^{\text{ZGN}}&=t_{\text{eff}}(\lambda)=t(3\lambda/4)\left[1+\alpha\left(1-\frac{1}{\sqrt{1+4\pi^2\lambda^2/3}}\right)\right],
\end{split}
\end{equation}
where $t(\lambda)$ is given by the Eq. (\ref{tlambda}). The previous equation does not depend upon the site. Substituting all the parameters and by using $\lambda=0.8$, the hopping parameter is $t_{\text{eft}}(\lambda=0.8)\approx0.19t_0$ for $\alpha=0.4$ and $t_{\text{eft}}(\lambda=0.8)\approx0.15t_0$ for $\alpha=0$. Thus, in these cases, the system has a ZGNs-like spectrum with hopping parameter $t_{\text{eff}}$. Although $t_{\text{eff}}$ depends upon $\alpha$, this effect is small and the spectrum at the points $\sigma=2/3,4/3$ is very narrow, as seen in Fig. \ref{sigzig}. For $\sigma$ near to that points the spectrum is highly fragmented, because the $\alpha$-dependent terms become important. They act as a perturbation potential, splitting the band structure, therefore, the spectrum is wider near to $\sigma=2/3,4/3$, as can be observed in Fig. \ref{zigEvskx} b).

Let us now discuss the transition observed in Fig. \ref{zigEvskx} as $\sigma$ goes from zero to $1$.

\subsubsection{Case $\sigma=1/4$}

For $\sigma=1/4$, the band structure and DOS are shown in Fig. \ref{zigEvskx} c). In this case, 
the unit cell has only four different atoms, with hopping parameters given by,
\begin{equation}
\begin{split}
t_1^{\text{ZGN}}&=t_1=t(\lambda/2)\left[1+\alpha\left(1-\frac{1}{\sqrt{1+\pi^2\lambda^2}}\right)\right]\\
t_2^{\text{ZGN}}&=t_2=t_0+t_0\alpha\left(1-\frac{1-\pi^2\lambda^2}{1+\pi^2\lambda^2}\right)\\
t_3^{\text{ZGN}}&=t_1\\
t_4^{\text{ZGN}}&=t(\lambda),
\end{split}
\end{equation}
where $t(\lambda)$ is defined in Eq. (\ref{tlambda}). The eigenenergies can be calculated exactly,
\begin{equation}
\begin{split}
E(k_k)=&\pm\left[t(\lambda)+t_2\right]\\
&\pm\sqrt{\left[t(\lambda)-t_2\right]^2+16t_1^2\cos^2{\left(\sqrt{3}k_x/2\right)}}.
\end{split}
\end{equation}
From the dispersion relation it can be seen that the system behaves as a linear chain with two different atoms and hopping parameter giving by $t_1$ and self energies $t(\lambda)$ and $t_2$. When $\sigma$ takes irrational values near to $\sigma=1/4$ the degeneration is broken and the DOS becomes spiky, as can be seen in Fig. \ref{zigEvskx} b.

Finally, the case $\sigma=1$ is displayed in Fig. \ref{zigEvskx}. Note that this case (Fig. \ref{zigEvskx} a) is the same as in Fig. \ref{armEvskx} c. Let us explain this feature. When $\sigma=1$, the hopping parameter takes two different values depending on the parity of $j$, if $j$ is odd, $t_j^{\text{ZGN}}=t_0$ whereas if it is even, $t_j^{\text{ZGN}}=t(\lambda=0.8)\approx0.05t_0$. So, the effective system is again made of { dimers} with intra-{ dimers} hopping integrals given by $t(\lambda=0.8)c(k_x)$ and $t_0$. Due to $t(\lambda=0.8)<<t_0$ the effective system can be seen as { dimers} with hopping parameter $t(\lambda=0.8)c(k_x)$ weakly coupled. Hence, the gap's size must be $2|2t(\lambda=0.8)-t_0|$, the main difference between AGNs is that here we have $2t(\lambda)$, due to $c(k_x)=2\cos{(\sqrt{3}k_x)/2}$. This prediction will be confirmed by calculating the eigenenergies analytically in the next subsection.
\subsubsection{Case $\sigma=1$}
We first calculate the hopping parameter,
\begin{equation}
t_j^{\text{ZGN}} = 
\begin{cases} 
t_0, & \mbox{if } j\mbox{ is even} \\ 
t(\lambda)=\exp{\left\{-\beta\left[\sqrt{1+4\lambda^2}-1\right]\right\}}, & \mbox{if } j\mbox{ is odd} .
\end{cases}
\end{equation}
Thus, the effective system just have four different atoms per unit cell, and the effective chain is made for { dimers} with hopping parameter $t_0c(k_x)$. The eigenenergies for this system as a function of $\lambda$ and $k_x$ are
\begin{equation}
E(k_x)=\pm t_0\pm2 t(\lambda)\cos{\left(\frac{\sqrt{3}}{2}k_x\right)}
\end{equation}
For confirming the gap's size predicted before, we calculate it from the previous equation, resulting in
\begin{equation}
\Delta_{\text{ZGN}}=2\left |2t(\lambda)-t_0\right|.
\end{equation}
In this case, a gap is opened for $\lambda\geq\lambda_C$, where 
\begin{equation}
\lambda_C=\frac{1}{2}\sqrt{\left(1+\frac{1}{\beta}\ln{2}\right)^2-1}\approx0.34.
\end{equation}
{ This minimal value of $\lambda_C$ 
for opening a gap at the Fermi level exceeds the elastic response of graphene and thus seems difficult to be observed experimentally.}

\section{Conclusions}
\label{conc}
Summarizing, we have analyzed the electronic properties of AGNs and ZGNs under uniaxial periodic ripples, using an exact mapping of the corresponding tight binding Hamiltonian into effective one dimensional chains. In particular, we studied uniaxial periodic ripples, finding complex spectra, gaps at the Fermi level, and flat bands for AGNs. All these features can be understood by looking at the effective system which are made of { dimers}. For instance, when $\sigma$ is commensurable with the characteristic period of the lattice the effective system behaves as { dimers} weakly coupled resulting, for $\lambda$ big, in flat bands for AGNs. However, when this is not the case, the reciprocal space becomes dense which results in a fractal spectrum.

 This work was  supported by Direcci\'on General de Asuntos del Personal Acad\'emico-Programa de Apoyos de Proyectos de Investigaci\'on e Innovaci\'on Tecnol\'ogica (DGAPA-PAPIIT) IN-$102513$ project, and by the Direcci\'on General de C\'omputo y de Tecnolog\'ias de la Informaci\'on y Comunicaci\'on (DGTIC)-NES center.\\ P. R.-T. acknowledges support from Consejo Nacional de Ciencia y Tecnolog\'ia (CONACYT) (Mexico). Gerardo Naumis thanks the DGAPA-PASPA  program for a sabbatical scholarship at the George Mason University in
 Fairfax, Virginia.

\bibliography{biblioArmChairGraphene}{}
\end{document}